# Modeling Climate Change Impact on Wind Power Resources Using Adaptive Neuro-Fuzzy Inference System

Narjes Nabipour [1], Amir Mosavi [2,3], Eva Hajnal [4], Laszlo Nadai [2], Shahab Shamshirband [5,6], Kwok-Wing Chau [7]

[1] Institute of Research and Development, Duy Tan University, Da Nang 550000, Vietnam;

[2] Kalman Kando Faculty of Electrical Engineering, Obuda University, 1034 Budapest, Hungary; amir.mosavi@kvk.uni-obuda.hu

[3] School of the Built Environment, Oxford Brookes University, Oxford OX3 0BP, UK; a.mosavi@brookes.ac.uk

[4] Alba Regia Technical Faculty, Obuda University, Szekesfehervar, Hungary

[5] Department for Management of Science and Technology Development, Ton Duc Thang University, Ho Chi Minh City, Vietnam

[6] Faculty of Information Technology, Ton Duc Thang University, Ho Chi Minh City, Vietnam

[7] Department of Civil and Environmental Engineering, Hong Kong Polytechnic University, Hung Hom, Hong Kong, China

**Abstract:** Climate change impacts and adaptations is subject to ongoing issues that attract the attention of many researchers. Insight into the wind power potential in an area and its probable variation due to climate change impacts can provide useful information for energy policymakers and strategists for sustainable development and management of the energy. In this study, spatial variation of wind power density at the turbine hub-height and its variability under future climatic scenarios are taken under consideration. An ANFIS based post-processing technique was employed to match the power outputs of the regional climate model with those obtained from the reference data. The near-surface wind data obtained from a regional climate model are employed to investigate climate change impacts on the wind power resources in the Caspian Sea. Subsequent to converting near-surface wind speed to turbine hub-height speed and computation of wind power density, the results have been investigated to reveal mean annual power, seasonal, and monthly variability for a 20-year period in the present (1981-2000) and in the future (2081-2100). The results of this study revealed that climate change does not affect the wind climate over the study area, remarkably. However, a small decrease was projected for future simulation revealing a slightly decrease in mean annual wind power in the future compared to historical simulations. Moreover, the results demonstrated strong variation in wind power in terms of temporal and spatial distribution when winter and summer have the highest values of power. The findings of this study indicated that the middle and northern parts of the Caspian Sea are placed with the highest values of wind power. However, the results of the post-processing technique using adaptive neuro-fuzzy inference system (ANFIS) model showed that the real potential of the wind power in the area is lower than those of projected from the regional climate model.

1. Introduction

There is an increasing demand for renewable energy to attenuate greenhouse gas impacts on human life, environment, and ecosystems. As a result of population growth and development in industry, global warming caused by human activities is an ongoing concern that requires further considerations. Cleaner production in energy supply, such as renewable energy is among suitable remedies to deal with this important issue. Wind power as the most common type of renewable energy and its spatial distribution throughout the world attracted the attention of energy policymakers and researchers to develop and build wind farms. Thus, many studies focused on wind power potential inland and sea areas. However, offshore and onshore wind resources are influenced less than land from topography, and therefore, it is expected to have a higher potential for this purpose. The Caspian Sea, a landlocked sea between Iran and Europe, plays an important role in the economy of the surrounding countries. In this regard, the rapid growth of the population in the region necessitates the application of green energy resources for the power supply. Thus, the evaluation of wind power resources in offshore and onshore areas can be considered as a tentative option to meet the increasing demand for energy in the future.

Global warming and climate change due to the increase in greenhouse gases are going to influence many atmospheric and oceanic phenomena directly and indirectly. Wind climate is an important atmospheric variable that has a significant effect on many different hydrologic and oceanic phenomena such as evapotranspiration, wind energy, gravity waves, etc. Therefore, assessment of future wind climate is a key step toward sustainable development, and it should be taken into consideration prior to the construction of wind farms and the installation of wind turbines. Several global circulation models (GCMs) have been developed by different institutes to project atmospheric and oceanic variables under different future climatic scenarios and assumptions. They are run using different numerical techniques and require high-performance computers. The globe is gridded with different spatial resolution and the basic equations are discrete using numerical approaches with desired time steps. Therefore, these models are time consuming and also require high-speed computers. However, these models have been run globally ignoring local topography and have coarse spatial resolutions. Therefore, their outputs should be localized according to regional features. Dynamical and statistical approaches are two common approaches applied for downscaling purposes. The latter approach employs statistical techniques to find a relationship between the predictand and predictor while neglects the physical and boundary conditions of the region. On the other hand, the dynamical approach derives regional climate models (RCMs) imposing boundary conditions and regional topography on the GCMs. In this regard, they are physically based models and provide higher consistency with the regional conditions.

Reyers, Pinto, & Moemken (2015) employed a statistical-dynamical downscaling for localization of GCM wind data over Europe. The approach was applied for near-surface wind speed at 10 m. The results revealed that the proposed method can be efficiently used for downscaling purposes. Moreover, several statistical techniques gaining different regression-based techniques and also Weibull based bias correction were applied by different researchers which downscale the wind speed (Sailor, Hu, Li, & Rosen, 2000; Shin, Jeong, & Heo, 2018; Winstral, Jonas, & Helbig, 2017). However, most of the previous studies focused on wind speed and fewer efforts were devoted to modifying wind direction in which it plays an important role in the efficiency of wind turbines and yields. Regarding statistical downscaling techniques, even though statistics of the data will be improved, however, the coarse resolution disregarding local topography will still be unresolved. Therefore, dynamical approaches can be employed to resolve this problem in order to achieve higher spatial resolution data. Coordinated regional climate downscaling experiment (CORDEX) provides regional downscaling for different domains throughout the world using different dynamical techniques. CORDEX outputs are been increasingly used for climate change impacts analysis for regional studies because of their higher resolutions (Dosio, 2016; Koenigk, Berg, & Döscher, 2015; Mariotti, Diallo, Coppola, & Giorgi, 2014). Formerly, machine learning models such as extreme learning machine (ELM), artificial neural network (ANN) and neuro-fuzzy systems have been successfully for wind speed forecasting (Panapakidis, Michailides, & Angelides, 2019; Yang and Chen, 2019), river flow and flood management (Cheng, Lin, Sun, & Chau, 2005; Fotovatikhah et al., 2018; Yaseen, Sulaiman, Deo, & Chau, 2018), predicting solar radiation (Beyaztas, Salih, Chau, Al-Ansari, & Yaseen, 2019), estimation of evaporation (Moazenzadeh, Mohammadi, Shamshirband, & Chau, 2018; Qasem et al., 2019; Salih et al., 2019), wind power estimation and extraction among their many other applications (Petković et al., 2014; Shamshirband et al., 2016). Khanali, Ahmadzadegan, Omid, Nasab, & Chau (2018) a genetic algorithm to get an optimized layout for wind farm turbines in Iran. The study revealed the importance of the installation layout of wind turbines on the power yield. However, to date, there is not a research study using these techniques for the post-processing computations of wind power. In this regard, it can be considered as a novel approach to employ soft computing techniques to post-process the results of the wind power for the future period.

Concerning climate change impacts on wind energy resources, Pryor and Barthelmie (2010) reviewed probable mechanisms in which global climate change variability can affect future wind energy resources and conditions. Rusu and Onea (2013) and Amirinia, Kamranzad, & Mafi (2017) evaluated wind and wave energy along the Caspian Sea. The results demonstrated adequate energy resources in the study area. Moreover, it was found that the northern part of the sea is more appropriate for energy extraction due to characterizing shallow water and a higher magnitude of the energy. Viviescas et al. (2019) and de Jong et al. (2019) explored future variability in renewable energies of wind and solar resources in Latin America and Brazil. The results showed that climate change might affect the energy resources negatively even though for some regions such as the northeast of Brazil, an increasing trend for both solar and wind energy resources was projected. Solaun and Cerdá (2020) investigated climate change impacts on wind power in wind farms of Spain indicating about 5% and 8% changes in future wind speed and power, respectively. Porté-Agel, Wu, & Chen (2013) investigated the effect of wind direction on turbine losses in a large wind field through a numerical framework. The results showed a strong dependency of the turbine efficiency on the wind direction (e.g., only 10 degrees change in wind direction may change the total power output by 43%). Chang, Chen, Tu, Yeh, & Wu (2015) reported higher wind energy density in the eastern half of the Taiwan Strait. However, they found that the wind power density will decrease slightly (about 3%) in the study area under future climatic conditions. Tobin et al. (2016) explored climate change impacts on wind power potential using CORDEX outputs to project future variability of the energy. According to this study, wind farm yields will undergo changes smaller than 5% in magnitude for most regions and models. Davy, Gnatiuk, Pettersson, & Bobylev (2018) investigated future changes in wind energy in the Black Sea using outputs of the Europe-CORDEX domain. It is noticed that both datasets including CORDEX and ERA-Interim simulations have been obtained using advanced numerical models in computational fluid mechanic simulations. Prior to exploring climate change impacts, they compared historical wind data of CORDEX with those of ERA-Interim reanalysis data indicating a good agreement between two data resources. Furthermore, the results revealed that the future scenarios would not cause significant negative impacts on the wind energy resources for the study area.

The main objective of this study is therefore to employ the Asian domain of CORDEX outputs covering the Caspian Sea as our case study for two future scenarios to investigate climate change impacts on the wind energy resources. In this regard, the historical data of near-surface wind are evaluated versus ERA-Interim reanalysis data as a reference dataset. Subsequently, the historical and future wind outputs obtained from the CORDEX are converted to wind speed at turbine hub-height and wind power density. Inter-annual, seasonal and intra-annual variability of wind power is projected for present and future periods to find climate change impacts. Finally, an ANFIS based post processing approach is applied to modify the computations of the climate model based on the reference data. In brief, the study gains of engineering applications of computer science (ANFIS as a machine learning technique) and data-driven model as well as outputs of advanced computational models to project wind power variation over the Caspian Sea. The study area, data, wind speed and power characteristics, a brief description of the post-processing technique and also the modeling procedures are described in section 2. The results of the study are discussed in detail in section 3. The conclusion will be presented in the last section of the manuscript.

## 2. Materials and methods
### 2.1. Case study

This study explores climate change impacts on wind power resources in the Caspian Sea. As a landlocked sea, the Caspian Sea is surrounded by five countries of Iran, Russia, Kazakhstan, Turkmenistan, and Azerbaijan. The study area is located between latitude 36.5°N and 47°N, longitude 47°E and 54°E (Figure 1), and it is bounded to Russia and Kazakhstan in the north, Iran to the south, Azerbaijan and Turkmenistan to the west and east. The bathymetry of the area is varied widely from very shallow water with less than 5m depth in the northern part to deep water exceeding 500 m in the middle and southern parts (Figure 1). With a water volume of about 78200 km$^3$ and with an area of 371000 km$^2$, the Caspian Sea is known as the largest inland water body (Likens et al., 2009). Winds are the most typical hydrometeorological feature of the Caspian Sea. They mainly blow from the north and east. During the winter, easterly winds created as a result of the Asiatic anti-cyclone are dominant while in the summer, northerly winds cause due to the Azores high-pressure blows frequently (Rusu and Onea, 2013). Such as a massive water body, there are a limited number of weather stations to record wind speed and direction. Therefore, reanalysis data such as ERA-Interim obtained from the European Center for Medium Weather Forecasting (ECMWF) are considered as common wind data resources when long term data are required. Details of the applied data are presented in the following subsection.

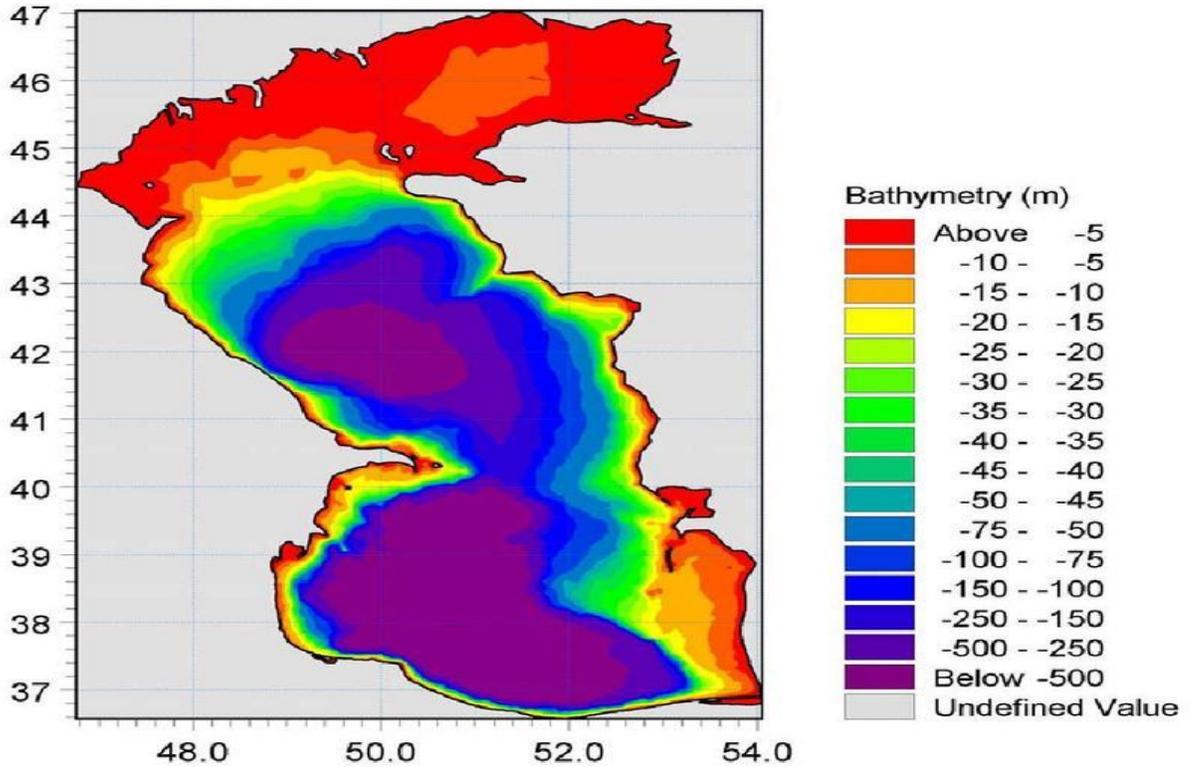

Figure 1. Depth variation of the study area, Caspian Sea (Allahdadi et al., 2004)

### 2.2. Data resources

Generally, the datasets used in this study for climate studies are obtained from CORDEX outputs which are regional downscaling of global circulation models. These outputs are derived by imposing boundary conditions from GCMs and run the model for a particular domain. Dealing with CORDEX, a domain is referred to as a region for which the regional downscaling is taking place. In this regard, these models have been developed for different regions (14 regions or domains) covering the whole globe altogether. The CORDEX outputs for each domain, regional climate model (RCM), usually provide the climatic data with higher spatial and temporal resolutions than the GCMs. The Caspian Sea domain is covered by the RCM of west Asia (WAS) with a spatial resolution of 0.44° in both latitude and longitude. Therefore, daily wind outputs if the RCM (WAS44i) have been obtained from the website for the study area. The wind components including zonal and meridional wind at 10 m from the surface are obtained. Prior to investigating climate change impacts, it is necessary to evaluate the consistency and accuracy of CORDEX outputs based on a reference datasets. Due to limited field observations and weather stations for such as a large water body, the ERA-Interim reanalysis wind data with a 6-hour interval are considered as reference data to assess the CORDEX for the historical period. There many GCMs can be used to derive RCMs while for the case of our study area and for domain WAS, only the CORDEX outputs obtained from GCM named MPI-ESM-LR (Max Plank Institute) are available. Figure 2 illustrates the mean annual wind speed for a 20 year period of 1981-2000 for ERA-Interim and the RCM.

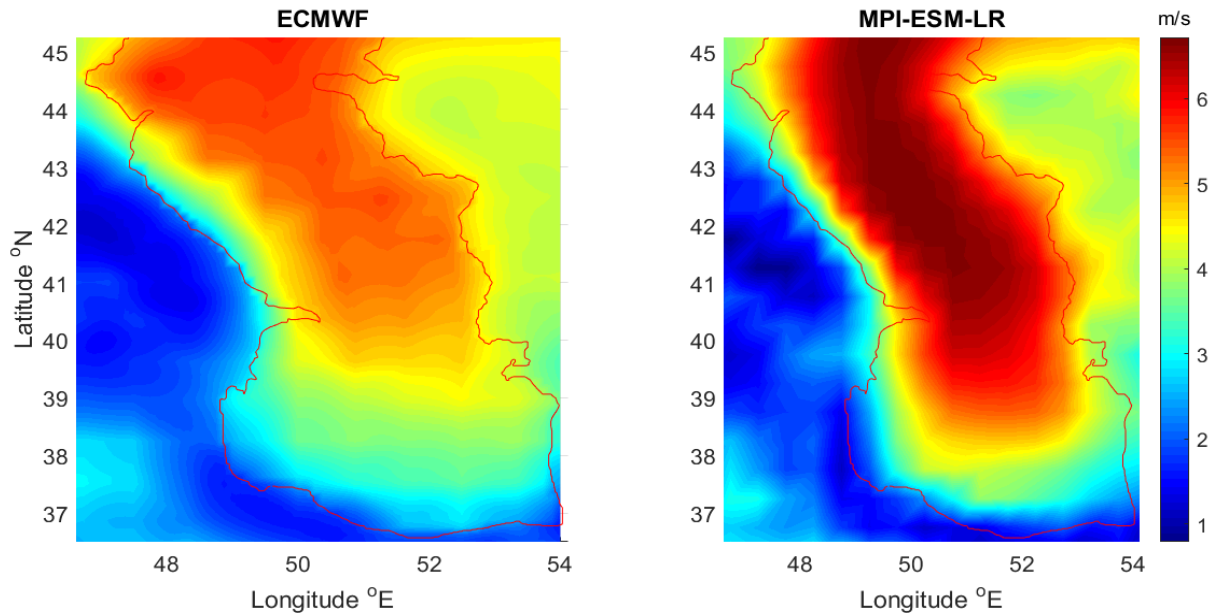

Figure 2. Mean annual wind speed for the historical period a) ECMWF and b) MPI wind simulations

According to Figure 2, it can be found that the regional climate model here means MPI-ESM-LR wind speed simulations overestimate wind speed over sea area while in the land surrounding the sea, the RCM slightly underestimates the wind speed. However, both data resources provide a similar trend, and there is a suitable consistency between two dataset simulations. Generally speaking, a higher gradient in wind data can be found for the RCM simulations than the corresponding values for the reference data (ECMWF). The average wind speeds over 20 years for the reference and the RCM simulations over the whole area illustrated in Figure 2 are 3.92 and 3.85 m/s, respectively. Therefore, a good agrees of similarity between reference and the RCM wind climate can be found for the historic period.

Subsequent to verifying the RCM wind data, the outputs are used for climate change impact studies. In this regard, historical and future wind outputs of the CODREX derived from MPI GCM for a 20 year period is used to compute wind power density is present (1981-2000) and future (2081-2100). To project the future distribution of wind power in the study area, two scenarios standing for two representative concentration pathways (RCPs) of RCP4.5 and RCP8.5 are taken under consideration. RCPs are greenhouse gas concentration trajectories adopted by the Intergovernmental Panel on Climate Change in its fifth assessment report. In this regard, four

RCPs of RCP2.6, RCP4.5, RCP6, and RCP8.5 indicating radiative forcing values in W/m² in the year 2100. These RCPs are developed according to a wide range of possible changes in future anthropogenic greenhouse gas emissions. RCP2.6 is known as an optimistic scenario, RCP4.5, and RCP6 as medium scenarios and RCP8.5 as the pessimistic scenario. However, this study employs CORDEX wind outputs for two RCPs of 4.5 and 8.5 which are commonly applied for climate change studies. More details about the future climate scenarios of the fifth assessment report can be found in Meinshausen et al. (2011).

### 2.3. Wind power computations

Generally, long term wind data are required for climate change impact studies in which they are usually available for near-surface at 10 m. However, wind power is mainly computed at turbine hub-height typically at 80 to 120 m. In this regard, near-surface wind data should be converted to turbine hub-height wind speed through an extrapolation technique. The extrapolation process is not an easy task because it is strongly dependent on the local conditions, such as roughness length (Stull, 2012). To do that, the power law is a commonly used expression that enables one to convert near-surface wind speed to any desired height following a limited number of local conditions (Eq.1).

$$W_z = \frac{u_*}{k} \ln \frac{z}{z_0} \qquad (1)$$

Where $W_z$ is the wind speed at turbine hub-height (z), $u_*$ is the shear velocity, $k$ is the Von Karman constant usually considered as 0.4, and $z_0$ is the roughness length. The sheer velocity itself is a function of the surface wind stress and the air density, which accurate calculation of these two parameters is a complex procedure due to their dependency on some other atmospheric variables such as pressure and temperature. Therefore, finding a suitable equation relating near-surface wind speeds to those of the turbine hub-height depending on fewer variables is of great interest. In this regard, the following equation based on only roughness length ($z_0$) and the reference height ($z_{ref}$) and the turbine height (z) is an appropriate alternative to determine the wind speed profile.

$$\frac{W_z}{W_{ref}} = \frac{\ln(\frac{z}{z_0})}{\ln(\frac{z_{ref}}{z_0})} \qquad (2)$$

Where all the variables in equation 2 can be easily determined except the roughness length that has its complexity. The surface roughness $z_0$ (in m) on onshore are constant while on offshore areas it is mainly dependent on wind waves. For the offshore regions, the surface roughness can be computed as:

$$\frac{z_0}{H_s} = 1200(\frac{H_s}{L_p})^{4.5} \qquad (3)$$

Where $H_s$ and $L_p$ are significant wave height and the wavelength of peak wave in m, respectively.

The roughness length for the calm state was measured by about 0.2 mm, and the average roughness for a blown sea was estimated at about 0.5 mm, as stated by Manwell, McGowan, & Rogers

(2010). Moreover, according to some other studies (Powell, Vickery, & Reinhold, 2003; Yu, Chowdhury, & Masters, 2008), it can be found that the roughness length for high winds and hurricanes can vary between 2 mm and 20 mm. However, for our case (Caspian Sea), there is no report indicating such high winds or hurricanes. Consequently, the wind state for the study area is considered a calm state. Furthermore, a study in the Caspian Sea revealed that the average roughness length is about 0.2 mm and the calm state conditions are dominated in the area (Amirinia, et al., 2017). Finally, having wind speed at the turbine hub-height, wind power density ($P_{wind}$) in terms of power (P) per unit area (A) or wind speed and air density can be computed as:

$$P_{wind} = \frac{P}{A} = \frac{1}{2}\rho_{air}W^3 \tag{4}$$

where $\rho_{air}$ is the air density hereafter equals to 1.225 kg/m³.

It is noticed that the main purpose of this study is to investigate climate change impacts on wind power in the Caspian Sea and detailed descriptions on roughness length, wind speed profile at turbine hub-height, and extrapolation of wind speed to the desired level are partially out of the scope of the manuscript. However, they can be found in the mentioned references.

### 2.4. ANFIS approach for post-processing

Data post-processing techniques are commonly used to modify or regionalize outputs of the numerical models to match better with those of the reference or real data. In this regard, regression-based models were traditionally employed to bring statistics of the computed data with the local measurements. However, these types of techniques mainly suffer from enough efficiency to catch nonlinear processes embedded in the phenomena. In the last decade, an adaptive neuro-fuzzy inference system (ANFIS) gaining both advantages of the artificial neural network and fuzzy systems attracted the attention of many researchers for time series forecasting and analysis. However, fewer efforts have been devoted to employing this approach for data post-processing purposes. As observed from Figure 2, the climate data obtained from CORDEX simulations show a remarkable overestimation compared to the reference data here means ERA-Interim. Therefore, even though this overestimation happens for both historical and future outputs of the climate model, for real evaluation and practical applications, a post-processing technique can be helpful to provide more accurate estimations of the wind power in the area. Since the climate models are gaining complicated numerical approach and they have passed many different validations, normality and other statistical tests, the historical and future outputs have been primarily investigated without any disturbance. Therefore, these data were first employed to explore climate change impacts on wind power potential and their temporal distributions. Afterward, when the climate change impacts have been revealed, an ANFIS based post-processing approach has been taken under consideration to bring the results in the reference data range. This is an important step toward sustainable development and future plan to extract wind energy resources.

In brief, an ANFIS model is combining transformed data through membership function with if-then fuzzy rules and an inference system to derive the desired results. Due to gaining from both features of neural networks and fuzzy logic, the ANFIS model uses the learning ability of neural networks to define the input-output relationship and construct the fuzzy rules by determining the input structure (Alizdeh, Joneyd, Motahhari, Ejlali, & Kiani, 2015). It is noted that this study is not to delve in detailed descriptions of the ANFIS model, and only the technique is applied from MATLAB library and further explanations including mathematical expressions of the method can be found in the relevant references (Alizadeh, Rajaee, & Motahari, 2016; Aqil, Kita, Yano, & Nishiyama, 2007). However, to illustrate the scheme briefly, a usual ANFIS is assumed to consist of 5 layers following each other. Figure 3 illustrates the schematic layout of a usual ANFIS consisting of 5 layers with two input variables of $x$ and $y$. The first layer is processing the input variables through a membership function. The input nodes in the first layer (represented in the figure as $A_1$, $A_2$, $B_1$, and $B_2$) are linked to the nodes in the second layer multiplying with the incoming nodes. In the third layer the normalized firing strength is calculated for each node (called weights $w_i$). The contribution of rules toward the output of the models are evaluated in this step. Finally, the last layer consisting a single node gives the outputs of the ANFIS model. In this study, first the ANFIS model was trained using historical wind power data of CORDEX and ECMWF as the model input and output, respectively. Subsequently, the model was fed by future data computed from CORDEX-MPI simulations to predict the modified wind power for future scenarios. Dealing with the ANFIS model, 4 Gaussian membership functions were employed used for data processing in the first layer.

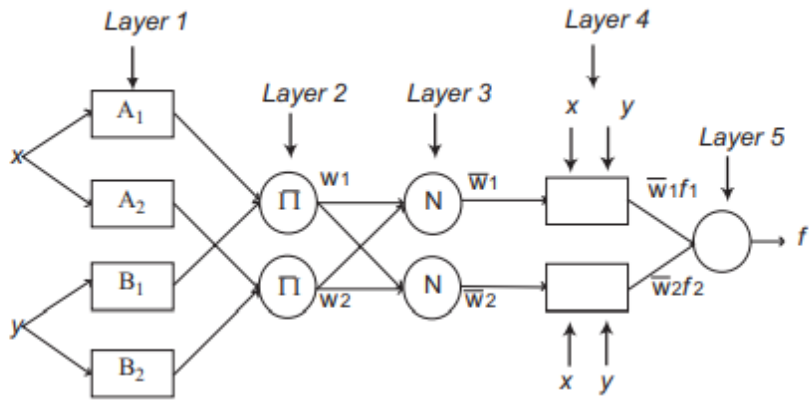

Figure 3. A schematic layout of ANFIS model (Aqil, et al., 2007)

### 2.5. Modeling procedures

To establish the projection of wind power and its variability under future climate change scenarios, historical and future simulations of near-surface wind was obtained from CORDEX. Prior to computing wind power from wind speed, the historical climate data of wind speed were evaluated against a reference data here means ECMWF (European Center for Medium-Range Weather Forecasts) reanalysis ERA-Interim data. Subsequent to the agreement between two wind resources, CORDEX simulations of wind data for historical (1981-2000) and future outputs for two climatic scenarios of RCP4.5 and RCP8.5 have been taken under consideration to determine climate change impacts. The CORDEX outputs were obtained for the global circulation model of MPI-EXM-LR developed by Max Plank Institute. Afterward, these data at 10 m were converted to the desired height named turbine hub-height through an extrapolation process. The converted wind speeds were employed to compute wind power density. By comparing the wind power for historical and future periods, the climate change impact was revealed. The results of climate change impacts in terms of annual, seasonal and intra-annual variability have been analyzed and mapped for the whole study area. In the end, a post-processing technique gaining the ability of the ANFIS model was constructed and employed to modify wind power computed from CORDEX wind simulations as it overestimates the wind speed compared to ECMWF wind data. In this regard, the mean annual values of wind power for the historical period of ECMWF and CORDEX have been reorganized in a single column matrix of input and outputs to train the ANFIS model. Furthermore, the ANFIS model was used to estimate the mean annual of wind power in different nodes of the study area for future scenarios in the testing stage. In brief, the main steps of the study can be summarized in Figure 4.

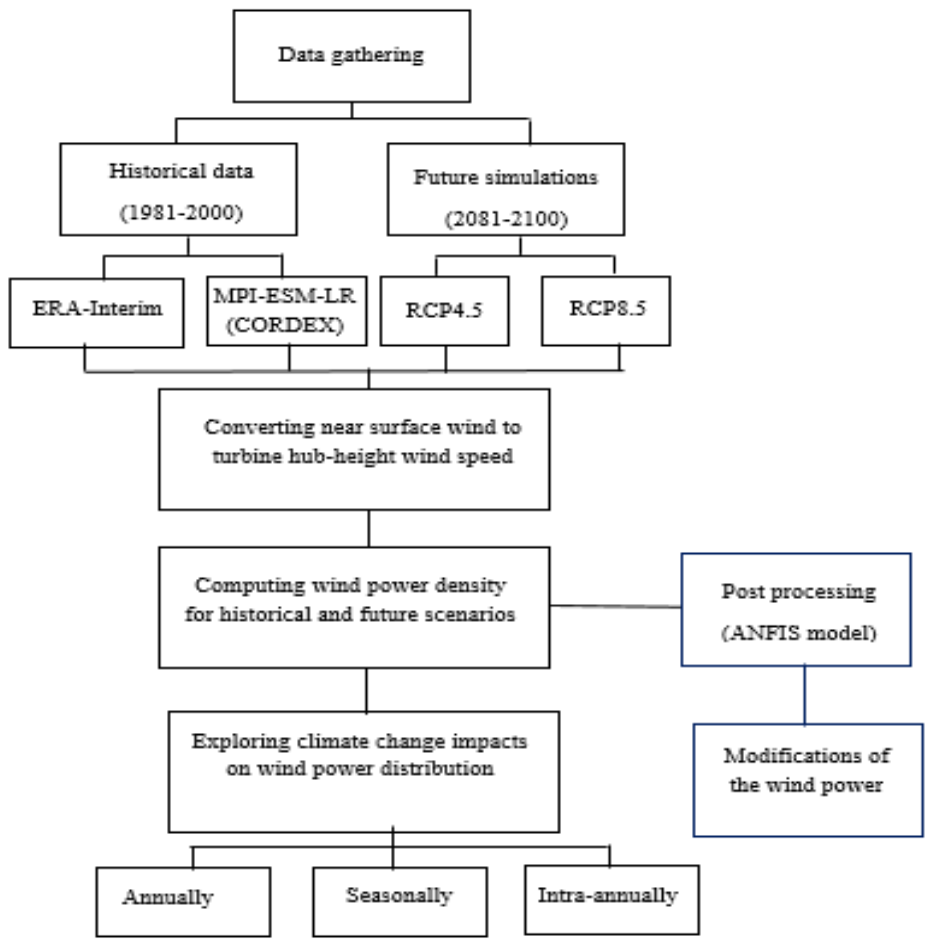

Figure 4. Schematic layout of the study

## 3. Results and discussion
### 3.1. Annual variability of wind power

To evaluate wind power distribution spatially and also to illustrate climate change impacts on future resources of wind power, the mean annual wind power distributed over the whole study area has been taken under consideration. To project future climate of wind power, CORDEX near-surface wind simulations for historical (1981-2000) and two future scenarios of RCP4.5 and RCP8.5 are employed for wind power computations at the turbine hub-height. Figure 5 illustrates the mean annual wind power over the study area for historical and future periods.

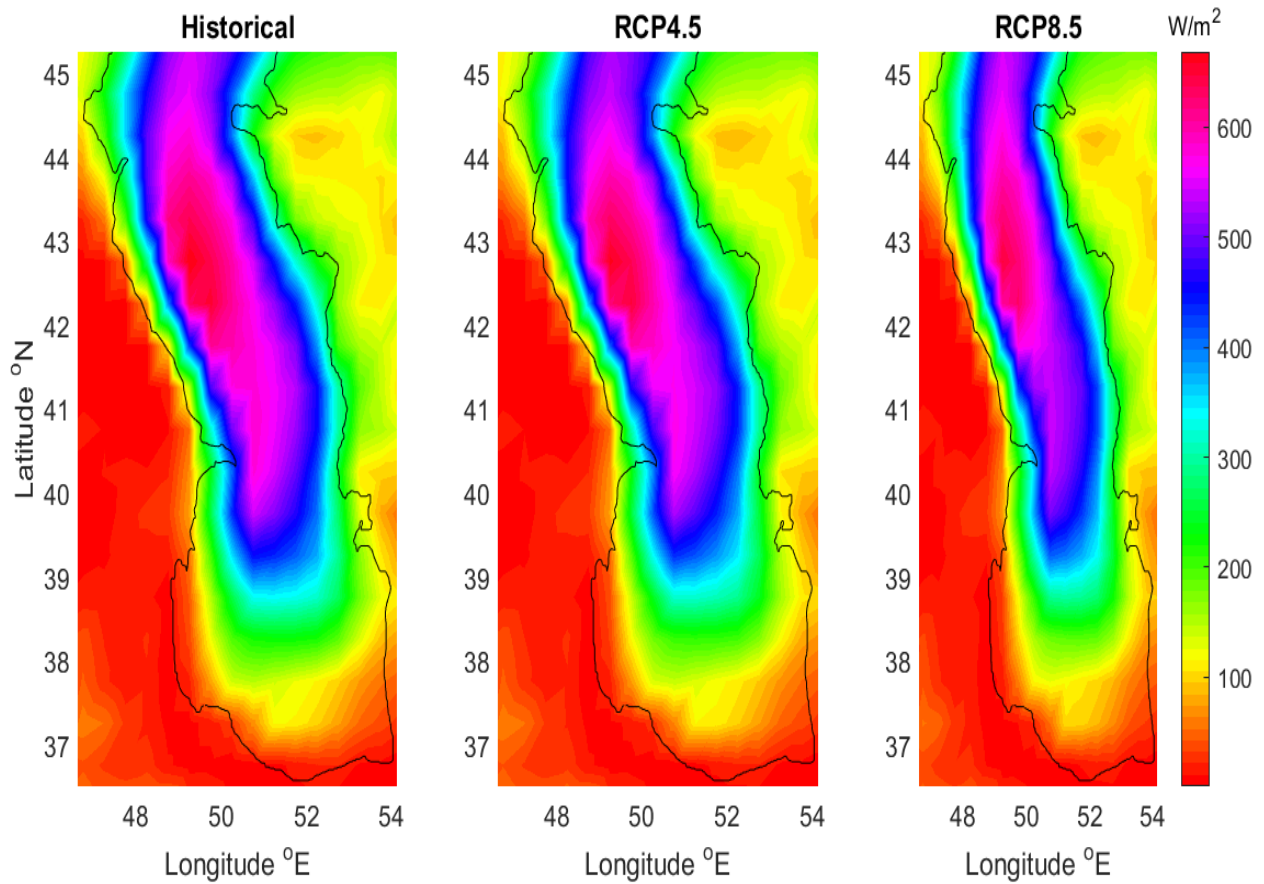

Figure 5. Projection of mean annual wind power over the study area for historical and future periods

Regarding Figure 5, it can be observed that the wind power over the study area has a quite varied distribution where for the middle part of the sea, the wind power is very high exceeding 600 W per unit area while for it southern part, the power decreases rapidly. Moreover, the gradient of wind power in offshore toward onshore or its surrounding lands strong. For example, on offshore, the wind power in the middle latitude and for the middle part of the sea is higher than 600 W/m$^2$ while it decreases as low as 100 W/m$^2$ in the eastern regions and the values are even less in western lands in the neighborhood of the sea. Generally, moving from the northern part of the sea toward the middle latitude, wind power increases and from the middle to lower latitude or southern part, the power decreases rapidly.

Considering climate change impacts on wind power distribution under the future scenarios, it can be seen that the historical projection of wind power has slightly higher values than the corresponding values for the future periods. Similarly, it can be found that the power for RCP4.5 are slightly stronger than those obtained for RCP8.5. However, the future changes in wind power compared to historical projection reveals that climate change does not affect future wind power significantly. This is an interesting point for future design and development of wind turbines in the study area. However, to illustrate mean annual future variations due to climate change impacts, relative changes of mean annual wind power for each scenario against historical projections have been computed, and the results are depicted in Figure 6.

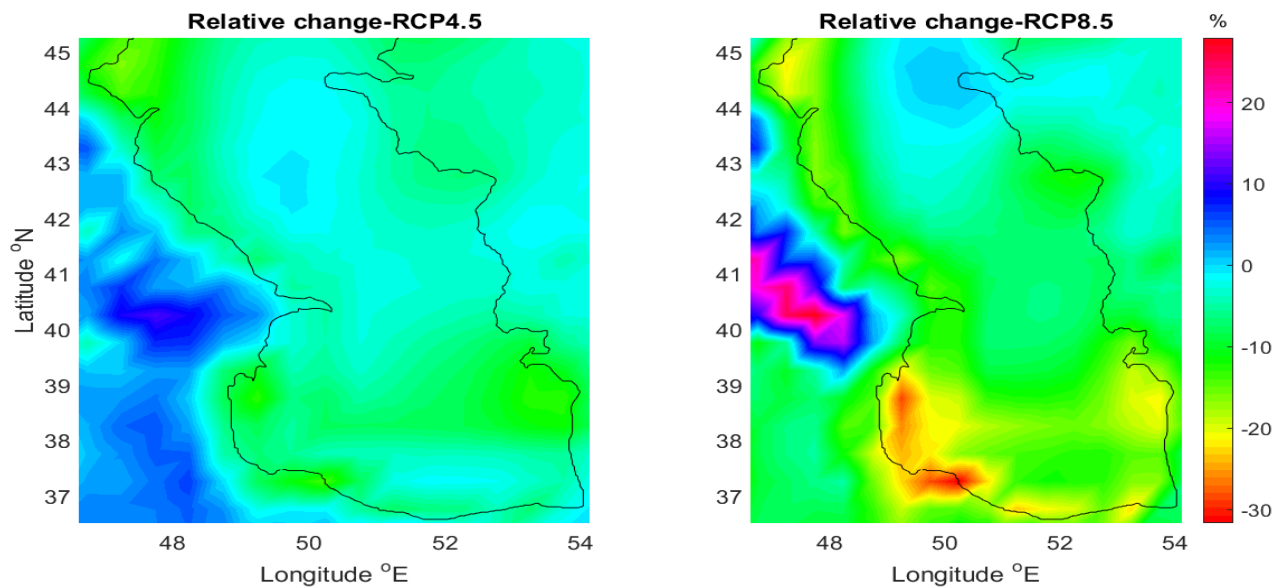

Figure 6. Relative changes in mean annual wind power for a) RCP4.5, and b) RCP8.5

As observed from Figure 6, the future changes in wind power are not remarkable for the major parts of the study area. Considering relative changes for RCP4.5, it can be derived that the future variation over the sea rarely exceeds 10%. Moreover, the negative sign of indicating by the colormap depicts the decrease in future wind power compared to historical values. Similarly, for the RCP8.5, a decrease over the whole sea is expected to happen in which the relative changes for RCP8.5 is a bit higher than RCP4.5. The change for RCP8.5 reaches 20 to 30% in the southeast of the sea. Therefore, the climate change impacts on the mean annual wind power for both scenarios provide a low decrease over the sea while the decrease is stronger in the southern part of the region where the wind power has lower values than the other regions. Therefore, the regions in the middle parts of the sea have a higher potential for power extraction. Despite the slightly decreasing trend in wind power projection over the sea, for another surrounding land especially in the eastern parts, a mild increase in wind power is projected. This increase is intensified for RCP8.5. However, further analyses are required toward reliable development and installation of wind turbine especially considering the temporal distribution of wind power and its future variation. The following subsections provide temporal projections of wind power in seasonal and monthly time scales for historical and future scenarios.

### 3.2. Seasonal variability of the power

Due to different temporal distribution over the seasons and also due to variable demand on wind energies as a result of different seasonal climate and difference in industrial and people activity, seasonal assessment of the power distribution can provide suitable information toward sustainable development. In this regard, wind power and its future variations for each season are computed separately, and the results are illustrated as the following figures. It is noticed that winter includes December, January, and February, and similarly spring is called for March, April, and May. Mean wind power for months June, July and August are integrated to compute mean wind power for summer. Finally, the wind power of September, October, and November are considered for the autumn.

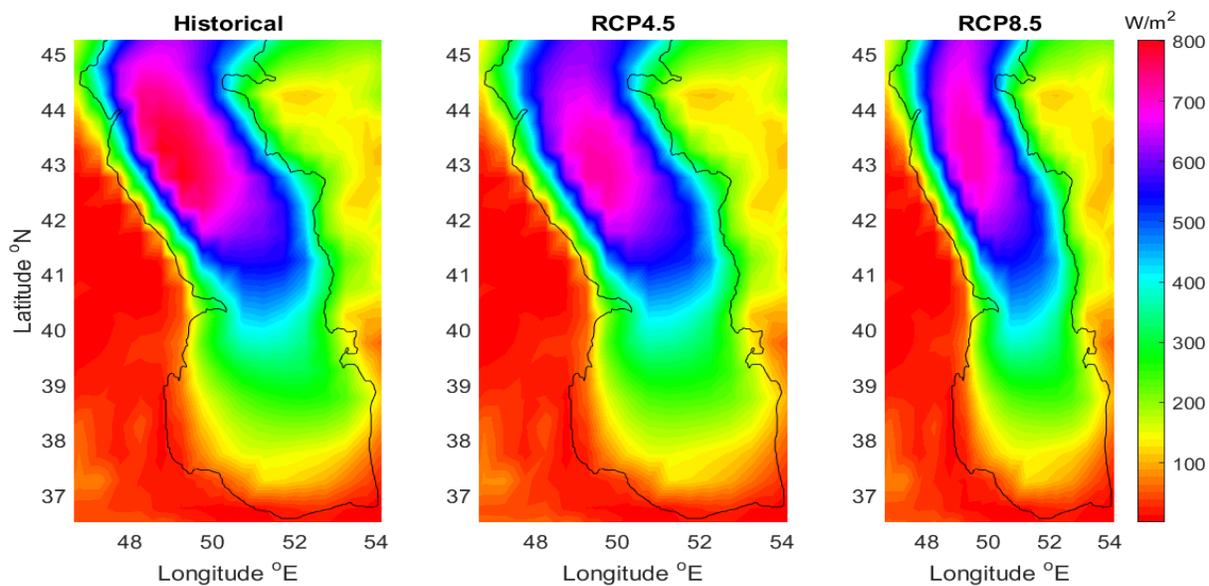

Figure 7. Wind power distribution in winter for historical and future simulations

Following Figure 7, the historical wind power has a bit higher value than the corresponding values obtained from future scenarios. The pink color in the left panel illustrating historical mean wind power for winter is more highlighted or is stronger than the future ones. Considering two scenarios, the RCP4.5 has a higher average than the other scenario, while for the offshore locations in the middle and northern part, the wind power distribution is higher for RCP8.5. The higher gradient of wind power for RCP8.5 than the alternative scenario balance these higher values in the middle and northern parts with those of lower ones in the eastern neighborhood of the sea. Comparing the distribution of winter with mean annual wind power (Figure 5), it can be derived that the trend in the winter is a little different from those illustrated in the annual plots. In the seasonal illustration for winter, the trend of peak values is extended from the high middle-latitude toward the northern parts, while for the annual, the trend is peak values are extended from the northern to the middle latitudes. Generally, the wind power in winter in average changes from 150 W/m² to 800 W/m² in the northern and middle parts of the sea. As the middle and northern parts have a relatively colder climate in winter and subsequently requiring more energy supply, therefore, wind power can be considered as a suitable option for energy supply as its distribution and power are remarkable for these areas. Moreover, future warming conditions do not change or decrease the wind power for these areas significantly. To proceed further, wind power distributions for the spring season are presented in Figure 8.

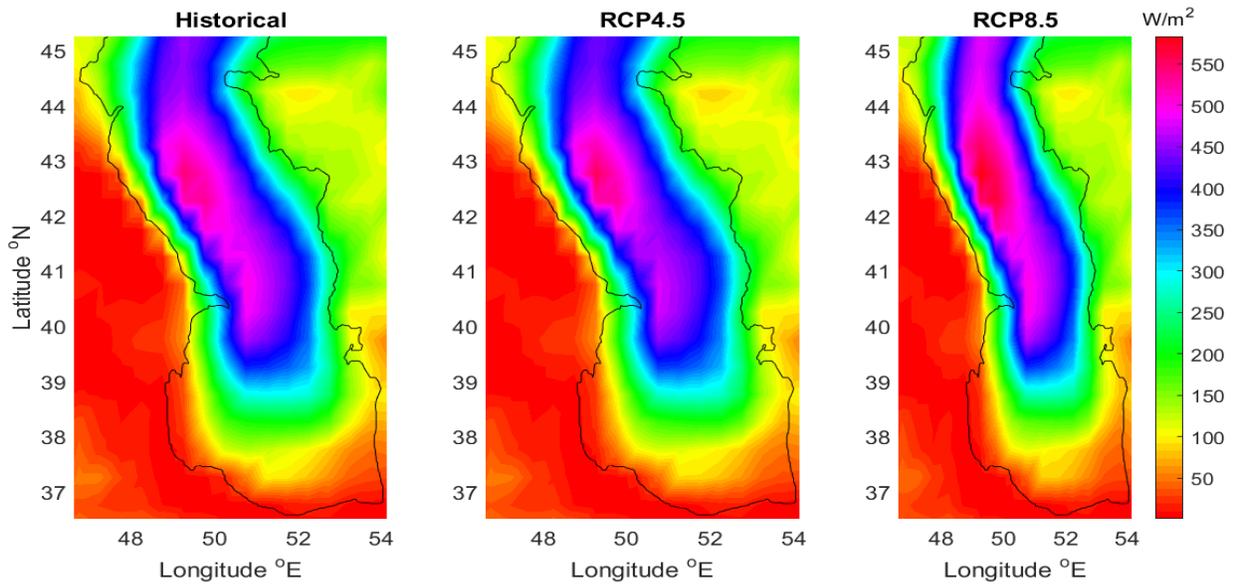

Figure 8. Wind power distribution in spring for historical and future simulations

Regarding Figure 8, the wind power in spring has a similar spatial trend with those of annual projections, while a different pattern for climate change impact projections can be found. However, the mean values of wind power in spring have relatively lower values than those of mean annual and also relatively lower values than winter. The mean power in spring is in a range of 50 to 550 W/m² indicating lower wind power in spring and governing a calm state over the sea. Considering future projections of wind power, the color indicating peak values of the power is more intensified for RCP8.5 than RCP4.5 and also than the historical corresponding values. Therefore, it can be found that the locations with the highest potential of wind power mainly in the middle latitudes will experience an increase even though the power is not expected to increase for the other regions. In a similar way, the results of wind power computations for historical and future projections in summer and autumn are illustrated in Figure 9 and Figure 10, respectively.

According to Figure 9, it is demonstrated that summer will experience relatively a strong decrease in wind power due to future climatic conditions. Moreover, it has an inverse spatially trend in the middle to northern latitude compared with those of annual, winter, and spring seasons. In other words, for summer, the strongest winds are blowing in low, middle latitudes while the general trends were found for high middle and northern parts. This is promising as the middle latitude, and lower has warmer summer and higher demand of energy supply than high middle latitudes and those of northern places. Considering the climate change impacts, future simulations of wind power provides less values than those of historical computations. Moreover, the RCP8.5 shows that the future decrease in summer is more intensified for RCP8.5 than RCP4.5. Generally speaking, it is demonstrated that wind power in summer may exceed 1000 W/m$^2$ which is much higher than the other seasons depicted earlier.

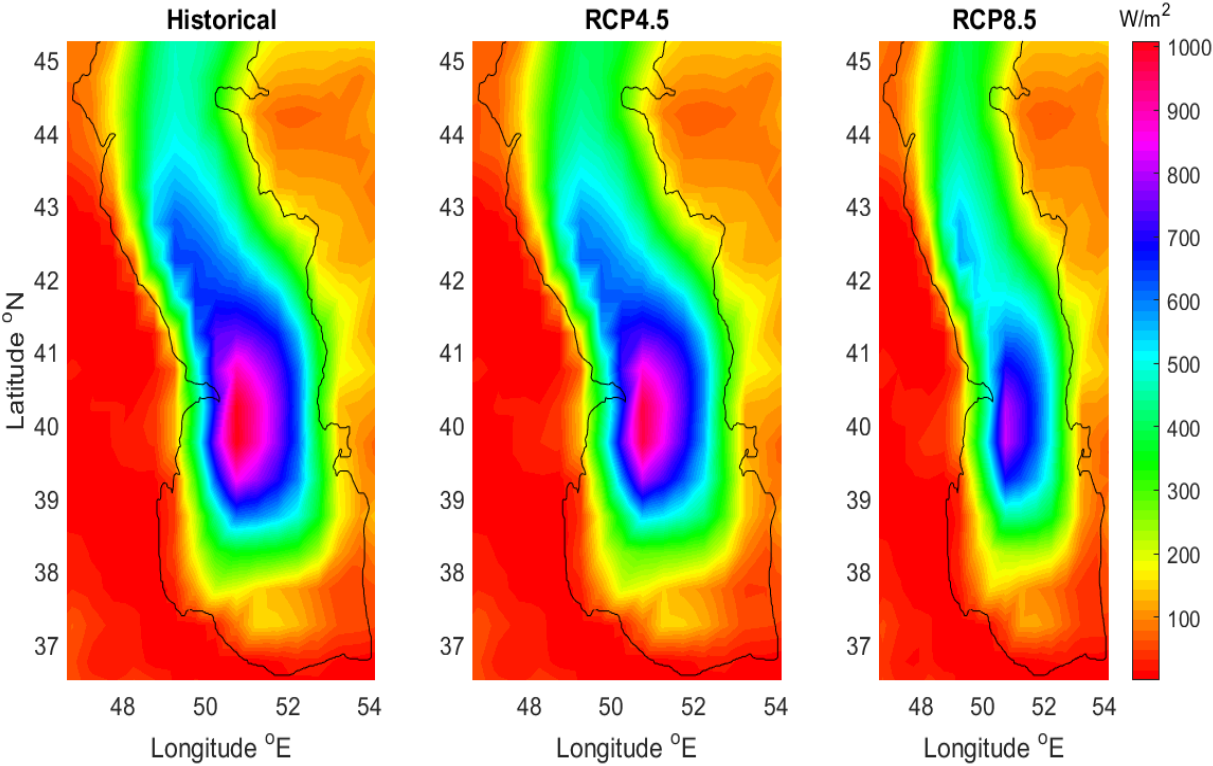

Figure 9. Wind power distribution in summer for historical and future simulations

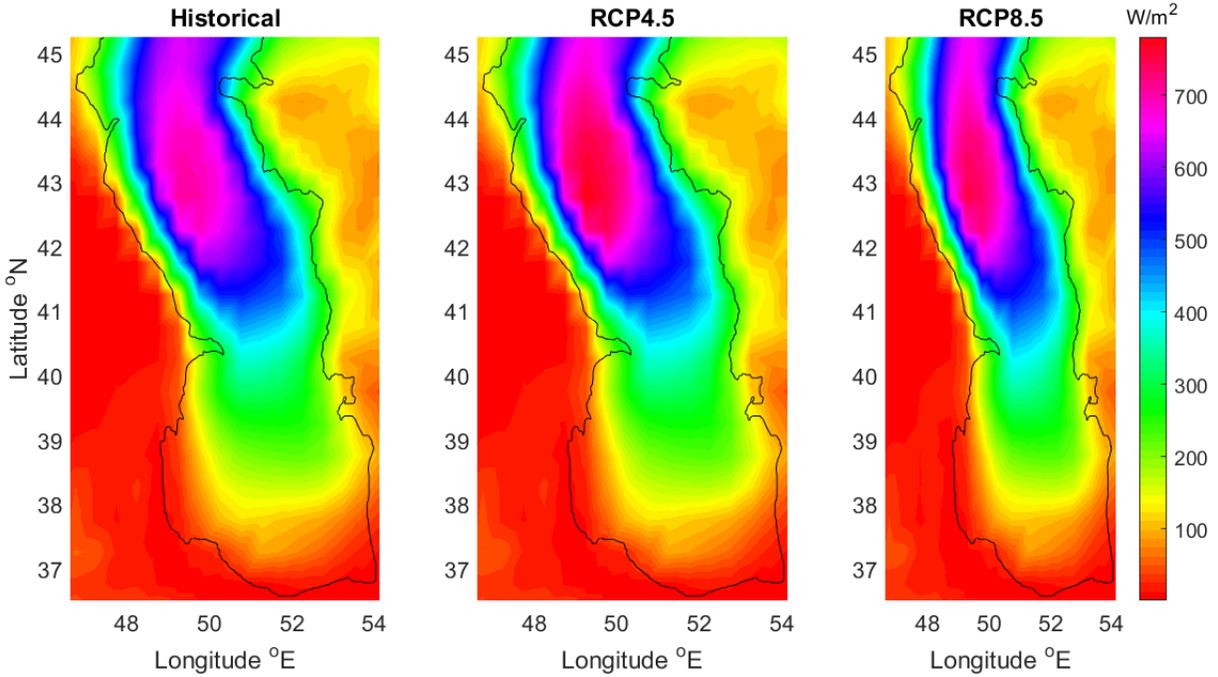

Figure 10. Wind power distribution in autumn for historical and future simulations

As illustrated in Figure 10, the climate change does not affect the wind power distribution over the study area remarkably. However, for the middle parts of sea representing the highest potential of the power, a slightly increase can be projected under future climatic conditions. Comparing the spatial distribution of the power for autumn with those of the other season and annual projections, it can be obtained the spatially distribution has a roughly similar trends with the annual one. Moreover, the computed values of wind power in autumn is closer to the mean annual power than those of computed for the other seasons. To provide more details of the wind power variation for different seasons and also for different scenarios, mean values of wind power averaged over the whole study area are presented in Table 1.

Table 1. Mean seasonal wind power averaged over the whole study area

| Season | Historical | RCP4.5 | RCP8.5 |
| --- | --- | --- | --- |
| Winter | 220 | 205 | 203 |
| Spring | 160 | 155 | 160 |
| Summer | 207 | 191.5 | 167 |
| Autumn | 182 | 186 | 183 |

The results of seasonal analysis given in Table 1 shows that historical projections have relatively higher power potential than the future simulations. However, the difference is negligible or is not valuable. Generally, the climate change impacts on seasonal distribution is different moving from one season to another. Moreover, the pattern is not monotonic while for winter and summer, the future simulations decrease remarkably in comparison with the historical values with higher rate of decrease for RCP8.5 than RCP4.5. For spring and autumn, the mean values of wind power do not different significantly under climate change scenarios. Also, seasonal projection for wind power revealed a high level of temporal distribution in which winter and summer are the most powerful seasons. On the other hand, spring season can be classified as the season with the lowest wind power potential.

### 3.3. Intra-annual distribution of the wind power

Intra-annual variability analysis is a common procedure to detect temporal distribution in climate change studies. It can provide more details of the variation of the power under future scenarios. Therefore, in this study, mean value of wind power for historical and future scenarios was average over the whole study area and the results are illustrated as Figure 11. In the figure, the vertical axis represents mean wind power in Watt per unit area and the horizontal axis represent month number in which starts from month number 1 as January continued to the last month of the year called December here denoted with the number 12. However, the months with number 12, 1 and 2 represent the winter season, numbers 3, 4 and 5 are spring, and the months for summer have numbers of 6, 7, and 8 and remained months are used for autumn calculations.

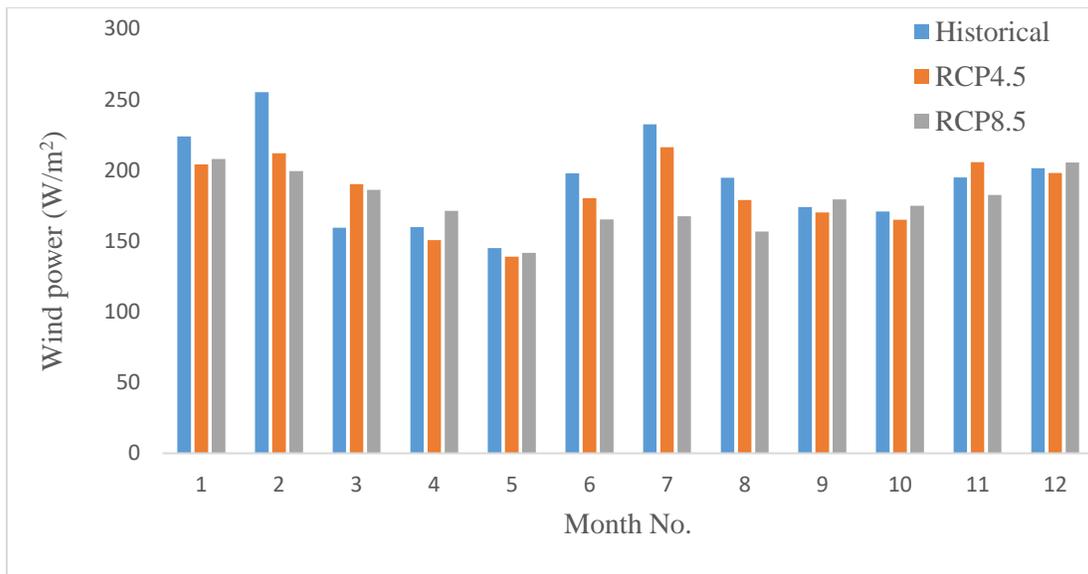

Figure 11. Intra-annual variability of wind power under future climatic scenarios

The monthly analysis of wind power depicted in Figure 11 presents high level of intra-annual variability when the months in winter and summer have the highest rate if changes due to climate change impacts. Overall, January and July are the months with the highest decrease under climate change impacts. On the other hands, May, September and October are the months that are less influenced under future climatic conditions. February, January, and July are the months with the highest potential of wind power and on the opposite, March and May have the lowest power amongst the other months. Generally, future simulations of wind power illustrates a lower values than the historical projections and also, simulations obtained from RCP8.5 has lower values than RCP4.5. However, the decrease is not consistent for all months and sporadic increase are observed for some months. Therefore, the climate change is expected to impose a slightly decreasing trend on future wind power resources even though this trend may not be regular for all the months or some inconsistency may be found.

### 3.4. ANFIS post-processing approach for data modification of CORDEX

To bring the CORDEX estimations of wind power to real world applications, consistent in magnitude with the reference data, the ANFIS model was developed in which the model used the CORDEX outputs as the predictor and the reference data as predictand. This process was carried out to train the model and also to predict and modify the future simulations of the CORDEX outputs applied for wind power computations. The predictor and predictand mean annual wind power over the whole study area were converted or reshaped to a single column matrix to meet the ANFIS model input requirement. After training and testing the model, the model outputs have been reorganized to their normal form to represent the power spatial distribution. The overall performance of the model during training process in terms of coefficient of determination was calculated as $R^2=0.59$. Figure 11 illustrates the CORDEX, ECMWF and the ANFIS outputs for the training stage (for the historical period).

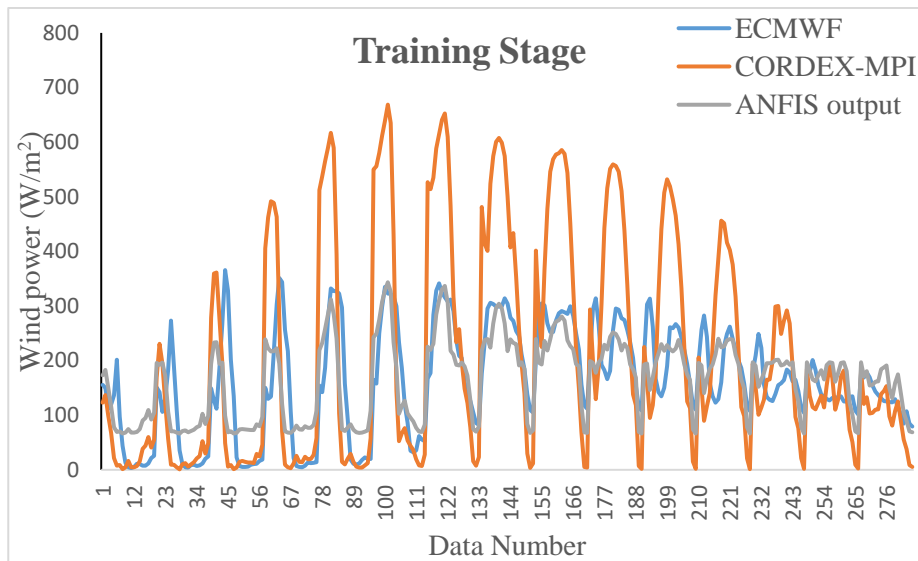

Figure 12. Results of the ANFIS model for the historical period

As observed from Figure 12, the climate model in the study area reveals that the CORDEX simulations provide higher values of wind power in regions with high potential of energy. On the other hand, for the regions with lower potential of wind energy, the CORDEX simulations provide lower values than the corresponding wind power of ECMWF. Therefore, a post processing can be deemed as an appropriate option to modify these values toward the real values. The ANFIS outputs illustrated in the figure indicate that the model in the training stage can be suitably used to bring the data in their reference data range. There is a good agree of similarity between the ANFIS outputs with those of the ECMWF based data. Subsequent to the model training, the tuned variables or weights have been used to estimate wind power for the future scenarios. In this regard, two models have been constructed separately to estimate wind power for RCP4.5 and RCP8.5. Figure 12 presents the results of the ANFIS models for the two future scenarios.

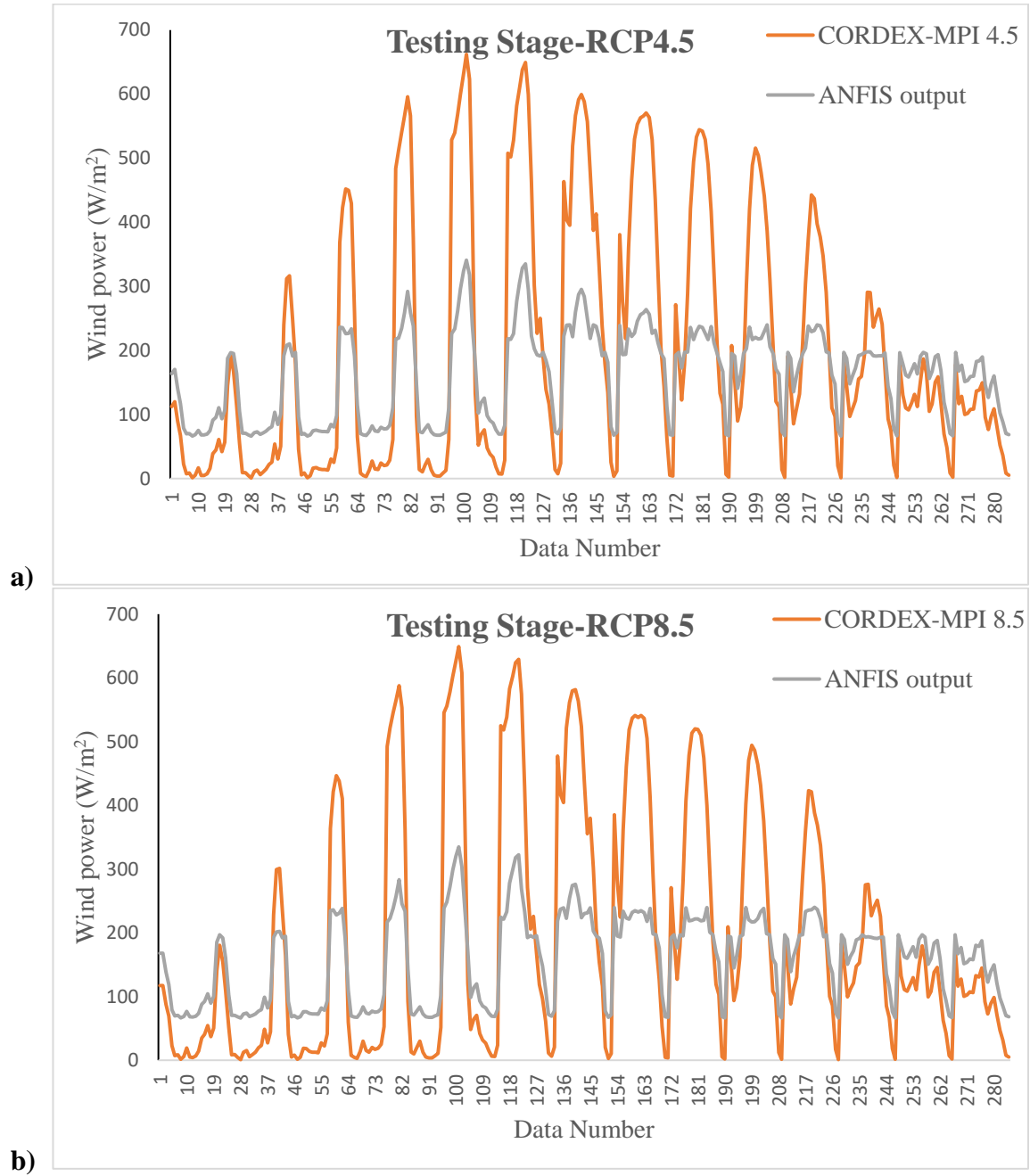

a)

b)

Figure 13. Results of the ANFIS model for a) RCP4.5, b) RCP8.5

According to Figure 13, the ANFIS model outputs for the future scenarios are remarkably lower than those of obtained from the CORDEX model directly. On the contrary, the model estimations for the places with lower potential are higher than those of the CORDEX based simulations. Therefore, the ANFIS model attenuate magnitude of the high potential spots and also augment magnitude of the locations with lower potential. However, outputs of ANFIS and CORDEX based computations have a homogeneous or similar trend but differing in the magnitude. The results of the testing period are in line with those of the training period in which the higher and lower estimations of CORDEX based wind power are decreased and increased in magnitude as the model outputs.

## 4. Conclusions

Global warming and climate change are subject of ongoing issues that their impacts are increasingly being considered on different environmental, atmospheric and aquatic ecosystems as well. On the other hand, renewable energy as a good remedy to attenuate greenhouse gas side effects and emissions have taken attention of many researchers. The main purpose of this study was to investigate climate change impacts of wind power resources in Caspian Sea. To do that, historical and future simulations (for a 20-year period) of near surface wind speed have been obtained from MPI-ESM-LR CRDEX outputs. The near surface wind speed was transferred to turbine hub-height of 120 m and the transformed wind speed was employed to compute wind power over the study area. The power distribution in terms of spatial and temporal variability for historical and two future climatic scenarios of RCP4.5 and RCP8.5 have been taken under consideration. Moreover, an ANFIS based model was developed to bring the statistics of the wind power obtained from CORDEX data to those of reference data.

Results of wind power projections revealed that the study area has a suitable potential for energy extraction as the mean annual wind power reached over 800 Watt per unit area. Spatial distribution of wind power over the study area present a quite variable distribution in which middle and northern parts of the sea has extremely higher potential for wind power compared to the southern stripe of the sea. Generally, future projections of wind power over the study shows a decreasing trend compared to the historical computations. However, the decrease has a strong temporal and spatial distributions in which for some special regions out of the sea the power may increase. Considering two investigated scenarios, it was derived that RCP8.5 provide higher gradient and rate of changes than RCP4.5. Results of relative change in wind power for future simulations against historical values proved this fact where the relative change for RCP4.5 was obtained in a range between 0 to -10%. However, corresponding values for RCP8.5 reaches -20 to -30% especially in the south eastern of the sea.

Considering temporal variability of wind power under future climatic conditions it can be derived that the future wind power in winter and summer decrease remarkably while for spring and autumn, the future changes are negligible. Moreover, it was found that the winter and summer are the most powerful seasons and spring has the lowest win power amongst the other seasons. The results of intra-annual variability demonstrated that February and July are the months with the highest rate of changes here means decrease for future scenarios. Simultaneously, they are considered along with January as the most powerful months. On the other hand, the months consisting of spring and autumn are expected to affect the least because of future climatic conditions. Among different months, May and October have the lowest rate of variations when future values are compared to those of historical values. Low values of wind power in spring may reflect calm state governing over the sea.

Generally, the results of this study are promising to develop and install wind turbines in the study area for energy extraction due to the suitable potential of wind power, especially in the middle and northern parts. Moreover, it was found that future climatic conditions do not change the wind power distribution in the areas significantly. More interestingly, the seasonal wind power resources give higher potential for winter and summer in which the energy demand for the area in these seasons. Therefore, there is a good consistency between the demand and supply of energy. In the end, it should be added that the real potential of the area is lower than those of computed from CORDEX simulations even though the modified values are still remarkable and implying the suitable potential of the area for wind power extraction. The findings of this study are beneficial for energy policymakers and strategists to plan for any development in the win power sector, considering the future variability of the power under future climatic scenarios. Future directions of the study can focus on uncertainties associated with the regional climate models and the efficiency of different climate models for the area.

## Acknowledgments:

We acknowledge the financial support of this research by the Hungarian State and the European Union under the EFOP-3.6.1-16-2016-00010 project.

Appendix.

Table A. List of abbreviations

| Abbreviation | Explanation |
| --- | --- |
| ANFIS | Adaptive Neuro-Fuzzy Inference System |
| CORDEX | Coordinated Regional climate Downscaling Experiment |

| ECMWF | European Center for Medium Weather Forecasting |
| --- | --- |
| GCM | Global Circulation Model |
| MPI | Max Plank Institute |
| RCM | Regional Climate Model |
| RCP | Representative Concentration Pathway |